\documentclass{aastex}
\usepackage{emulateapj5}
\usepackage{apjfonts}

\submitted{Accepted by the Astronomical Journal 2003 March 18}

\shortauthors{Boboltz et al.}
\shorttitle{Radio Star Positions and Proper Motions}

\begin{document}

\title{Astrometric Positions and Proper Motions of 19 Radio Stars} 
\author{D. A. Boboltz\altaffilmark{1}, 
A. L. Fey\altaffilmark{1}, 
K. J. Johnston\altaffilmark{1},
M. J Claussen\altaffilmark{2},
C. de Vegt\altaffilmark{3}\altaffilmark{4},
N. Zacharias\altaffilmark{1}, and R. A. Gaume\altaffilmark{1}}

\altaffiltext{1}{U.S. Naval Observatory,
3450 Massachusetts Ave., NW, Washington, DC 20392-5420}
\altaffiltext{2}{National Radio Astronomy Observatory,
P.O. Box O, Socorro, NM 87801}
\altaffiltext{3}{Hamburger Sternwarte, Universitat Hamburg, Gojenbergsweg 112, 
D-201029 Hamburg, Germany}
\altaffiltext{4}{Deceased, 2002 July 12.}

\begin{abstract}  

We have used the Very Large Array, linked with the Pie Town Very Long Baseline
Array antenna, to determine astrometric positions of 19 radio stars in the 
International Celestial Reference Frame (ICRF).  The positions of these stars 
were directly linked to the positions of distant quasars through phase referencing 
observations.  The positions of the ICRF quasars are known to 0.25~mas, 
thus providing an absolute reference at the angular resolution of our 
radio observations.  Average values for the errors in our derived positions for 
all sources were 13~mas and 16~mas in $\alpha \cos\delta$ and $\delta$ respectively, 
with accuracies approaching 1--2~mas for some of the stars observed.  
Differences between the ICRF positions of the 38 quasars, and those 
measured from our observations showed no systematic offsets, with mean values of 
$-$0.3~mas in $\alpha \cos\delta$ and $-$1.0~mas in $\delta$.  Standard deviations 
of the quasar position differences of 17~mas and 11~mas in $\alpha \cos\delta$ 
and $\delta$ respectively, are consistent with the mean position 
errors determined for the stars.  Our measured positions were combined with 
previous Very Large Array measurements taken from 1978--1995 to determine 
the proper motions of 15 of the stars in our list.  With mean errors of 
$\approx$1.6~mas~yr$^{-1}$, the accuracies of our proper motions approach those 
derived from Hipparcos, and for a few of the stars in our program, are better 
than the Hipparcos values.  Comparing the positions of our radio stars with 
the Hipparcos catalog, we find that at the epoch of our observations, the 
the two frames are aligned to within formal errors of approximately 3~mas.  
This result confirms that the Hipparcos frame is inertial at the expected 
level.

\end{abstract}


\section{INTRODUCTION}

The current realization of the International Celestial Reference Frame (ICRF) is defined by 
the positions of 212 extragalactic objects derived from Very Long Baseline Interferometry (VLBI)
observations \citep{MA:98}.  This VLBI realization of the ICRF is currently the International 
Astronomical Union (IAU) sanctioned fundamental astronomical reference frame. 
At optical wavelengths, the Hipparcos catalog \citep{PERRYMAN:97} now serves as 
the primary realization of the extragalactic frame.  The link between the Hipparcos 
catalog and the ICRF was accomplished through a variety of ground-based and space-based 
efforts \citep{KOVAL:97} with the highest weight given to VLBI observations of 12
radio stars by \cite{LPJPRTRG:99}.  The standard error of the alignment was estimated 
to be 0.6~mas at epoch 1991.25, with an estimated error in the system rotation of 
0.25~mas~yr$^{-1}$ per axis.  Thus at the epoch of our observations (2000.94) the alignment
of the Hipparcos frame and the ICRF had a formal error of approximately 2.5~mas.  Due to
errors in the proper motions, the random position errors of individual Hipparcos stars
increased from $\sim$1~mas in 1991 to $\sim$10~mas at the time of our observations. 
Upcoming astrometric satellite missions such as SIM and GAIA will likely define frames 
with internal accuracies that are better than the extragalactic VLBI frame by an order 
of magnitude, and these frames may define the next generation ICRF. 

In this paper, we present our observations of 19 radio stars 
using the Very Large Array (VLA) in A configuration linked by fiber optic transmission 
line to the Very Long Baseline Array (VLBA) antenna located in Pie Town, New Mexico. 
Both the VLA and VLBA are maintained and operated by the National Radio Astronomy 
Observatory (NRAO)\footnote{The National Radio Astronomy Observatory is a facility 
of the National Science Foundation operated under cooperative agreement by Associated 
Universities, Inc.}. The VLA plus Pie Town (VLA+PT) link \citep{CBSU:99} is a valuable tool for 
radio star astrometry because it provides the high sensitivity of the VLA 
with nearly twice the resolution of the VLA A-configuration alone for high declination sources.  
In sections 2 and 3 we describe the observations, reduction of the data,
and the methodology used in the determination of source positions and associated errors.  
The observations described here represent a continuation of a 
long-term program (since 1978) to obtain accurate astrometric radio 
positions, parallaxes, and proper motions for $\sim$50 radio stars, which can be used to 
connect the current ICRF to future astrometric satellite reference frames. 
In section 5 we combine our VLA+PT positions with previous positions derived 
from VLA data collected from 1978 through 1995 \citep{JWFD:85, JDG:03}.  We derive 
updated estimates of the proper motions for 15 of the stars observed in both programs, 
and compare these proper motions with the corresponding Hipparcos values. 

\section{OBSERVATIONS AND REDUCTION \label{OBS}}

The VLA+PT X-band observations occurred over a 24-hour period beginning 
2000 December 10 at 06:30 LST.  The data were recorded in dual circular
polarization using two adjacent 50-MHz intermediate frequencies (IF's) 
centered on rest frequencies of 8460.1 MHz and 8510.1 MHz respectively.
The distribution on the sky of the 19 radio stars observed is shown 
in Figure \ref{AITOFF}.  For each star, two nearby ICRF reference 
sources were observed for phase calibration (see Table \ref{SOURCES}). 
The input positions for the stars were Hipparcos values updated to the epoch
of our observations using the Hipparcos proper motions and parallaxes.  Input 
positions for the ICRF reference sources were those given in \cite{IERS:99}. 
Because radio stars are generally weak (on the order of a few mJy) and because the 
goal of our observations was astrometry, negating the use of phase self calibration, we used 
the fast-switching technique \citep{CH:97} to observe the star and its primary phase 
calibrator in an attempt to mitigate phase fluctuations due to the atmosphere/ionosphere. 
Each fast-switched scan was bracketed by two short scans of a second ICRF source which 
was used as a phase calibration back-up and as a check for the accuracy of the 
position estimation.  Over the 24-hour experiment, three observations of each 
radio star were conducted at three different hour angles in order to maximize 
the $uv$ coverage.  For the discussion presented here, an observation is defined 
as consisting of a short 1.5 minute scan on the secondary phase calibrator, 
followed by 17.6 minutes of fast switching between the star and the 
primary phase calibrator with a 2.5 minute cycle time (100 seconds on the star and 
50 seconds on the calibrator), followed by a final 1.5 minute scan on the secondary phase 
calibrator. In addition, two 5 minute scans were recorded on the sources 
3C48 and 3C147 for use in the absolute flux density calibration.
 
Data were reduced using the standard routines within the Astronomical 
Image Processing System (AIPS).  The absolute flux density scale was established 
using the values calculated by AIPS for 3C48 and 3C147 (3.22~Jy and 4.81~Jy respectively)
with the proper $uv$ restrictions applied.  For each star, two calibrated 
$uv$ data sets were generated.  For the first, the phase calibration was accomplished 
through transfer of the phases of the primary (fast-switched) reference source.  
A second $uv$ data set was generated for each star by applying the 
phases of the second ICRF calibrator (not fast-switched) bracketing the corresponding star.  
Each ICRF reference source was phase calibrated with the other reference source 
observed, i.e. the primary was calibrated using the secondary and the secondary 
calibrated using the primary.  At no time was self-calibration performed on any of the 
data.  From the $uv$ data sets (two per star and one per calibrator), a set of five 
images were generated for each source: one CLEANed image including all observations, 
one dirty image with all observations, and three dirty images one per observation 
at the three different hour angles.  Synthesized
beam sizes ranged from approximately $0''.35\times0''.15$ for a source at a declination of 
$-28^{\circ}$ to $0''.17\times0''.14$ for a source at a declination of 
$+45^{\circ}$.  The images produced have 512$\times$512 pixels with a 
pixel size of $0''.015$.  For the CLEANed images 100 iterations were used.  
A two-dimensional (2-D) Gaussian function was fit to the peak flux in each image using 
the AIPS task JMFIT.  The results of these fits were used in the derivation of source 
positions described in the next section. 

\section{POSITION DETERMINATION \label{POS}}

Final estimation of the star and calibrator positions was performed outside of AIPS 
using the results of the 2-D Gaussian fits to the various images.  Peak and RMS flux 
densities for each source were derived from the fits to the CLEANed images.  
To avoid any possible position shifts due to CLEANing
of the images, positions in right ascension and declination were determined
from fits to the dirty images produced from the combined data for each source. 
A later comparison showed less than 1--2~mas differences between the positions
derived from the dirty and CLEANed images in all but one case.  The star 
B Per, the lowest flux density source observed,  had a difference in right ascension 
of 5 mas, which was still well within the position error reported for this source.
Table \ref{POSITIONS} lists the positions and flux densities determined from the
fits to the images of the 19 stars and their associated calibrator sources.  
Other than the tropospheric delay correction made by the VLA on-line 
system, which assumes a plane parallel slab model based on pressure, 
temperature and relative humidity measurements made at the VLA, 
no other correction was made for tropospheric/ionospheric effects.  
Figure~\ref{ELEVERROR} plots the position formal errors (as determined by JMFIT)
in $\alpha \cos\delta$ and $\delta$ as a function of source 
elevation for all calibrator and target source 
observations.  As seen in the figure, there is a small rise in position error 
for sources observed at elevations below about $25^{\circ}$.  We made no attempt 
to model this phenomenon, but we did attempt to estimate contributions to the 
position errors due to tropospheric/ionospheric effects, as discussed below.

Documentation for the AIPS task JMFIT states that the errors in the position 
estimates should be regarded as tentative.  Taking a conservative approach to error 
determination, we computed a root-sum-square combination of two separate error 
estimates.  The first estimate, which is associated with the fitting of a 2-D Gaussian function 
to an image peak, is given by $\sigma \approx \theta_{beam}/ 2 SNR$, where SNR 
is the signal-to-noise ratio in the CLEANed image and $\theta_{beam}$ 
represents the geometric mean of the synthesized beam.  The second estimate, 
which provides a measure of the uncertainty due to the changing troposphere/ionosphere, 
is a weighted root-mean-square (WRMS) position error computed in the following manner.  
For each individual observation, positions and associated errors in right ascension 
and declination were determined from a 2-D fit to the dirty images at three different
hour angles.  A WRMS position uncertainty was then computed for each 
source, with the individual position errors reported by JMFIT used as weights.
The final errors reported in Table \ref{POSITIONS} are the combined root-sum-square of 
the two separate estimates.  For most of the stars in our observations, the final
reported errors were dominated by the WRMS uncertainties in the positions,
with the errors due to Gaussian fitting being a small contributor.  

For three of the stars in Table \ref{POSITIONS}, HD50896-N, RS CVn, and HR5110, 
phase stability was found to be better using the secondary ICRF calibrator rather 
than the primary (fast-switched) calibrator as the phase reference source.  
Therefore, the positions and errors reported for these stars are from the data which were 
phase-calibrated using the secondary (not fast-switched) reference source.  Mean values 
for the position errors for all 19 stars are 10~msec in right 
ascension, $\alpha$, (13~mas in $\alpha \cos\delta$)
and 16~mas in declination, $\delta$.  Upon close inspection of Table \ref{POSITIONS}, 
it is obvious that the errors associated with the star HD50896-N and its two 
calibrator sources are substantially higher than errors for all of the other 
sources.  This star was the second lowest declination star in our list, and 
its primary phase calibrator was the lowest declination source observed.  
In addition, one of the scans on this source was taken at a very low elevation 
(star at $\approx$16$^{\circ}$ and primary phase calibrator at 
$\approx$11$^{\circ}$).  Disregarding HD50896-N and its associated calibrator sources, 
decreases the mean errors in the positions for all sources to 12~mas in both 
$\alpha \cos\delta$ and $\delta$.

The lower limit on the accuracy of the positions for the best ICRF calibrators is
0.25~mas \citep{MA:98}, well below the precision obtained from our VLA+PT 
measurements discussed above.  The ICRF coordinates thus provide representative 
reference positions with which to compare our VLA+PT positions.  Figures \ref{ICRFVLARA} and 
\ref{ICRFVLADEC} display the results of such a comparison.  Differences between the 
ICRF positions and the VLA+PT positions for the calibrator sources are plotted as a 
function of source right ascension in Figure \ref{ICRFVLARA} and as a function 
of source declination in Figure \ref{ICRFVLADEC}.  Error bars in the two figures 
are those derived for our VLA+PT measurements reported in Table \ref{POSITIONS}. 
Neither figure shows a clear dependence of position offset on source 
right ascension or declination.  The means of the differences in $\alpha \cos\delta$ and 
$\delta$ are $-$0.3~mas and $-$1.0~mas respectively, indicating that systematic 
effects are negligible.  The standard deviations of the position differences 
are 17~mas and 11~mas in $\alpha \cos\delta$ and $\delta$  respectively.  These 
values are roughly equivalent to the mean position errors for the stars derived above.  
If we again disregard the two calibrator sources associated with the 
star HD50896-N then the standard deviations of the differences fall to 10~mas in 
both $\alpha \cos\delta$ and $\delta$.  

\section{RADIO/OPTICAL FRAME ALIGNMENT \label{FRAMES}}

In addition to the comparison of calibrator positions, we also 
compared the positions of the 19 stars as derived from our VLA+PT observations
with the corresponding Hipparcos positions updated to the epoch 
of our observations, Julian Day 2451889.  Figures \ref{HIPVLARA} 
and \ref{HIPVLADEC} plot the position differences for the radio stars as a function 
of source right ascension (Figure \ref{HIPVLARA}) and as a function of source declination 
(Figure \ref{HIPVLADEC}).  Error bars are those derived from our VLA+PT observations.
From Figures \ref{HIPVLARA} and \ref{HIPVLADEC}, it is apparent
that roughly half of the positions derived from our VLA+PT data do not agree with the 
Hipparcos positions to within the uncertainties in our measurements.  The most obvious 
disagreement is for the star UX Ari, for which the differences in $\alpha \cos\delta$ 
and $\delta$ are 84~mas and 42~mas respectively.  The large offsets for 
UX Ari are real, and are discussed in section 6 below.

Optical (Hipparcos) minus radio (this paper) position differences 
($\Delta \alpha \cos \delta, \Delta \delta$) were calculated for the 18 stars on 
our list (excluding UX Ari) at our 2000.94 epoch.  The three rotation angles between 
the optical and radio frames were determined from these data using a weighted 
least-squares adjustment (see Table \ref{ROT_TAB}).  For the radio positions,
we used the errors reported in Table 2.  For the optical positions we 
used the Hipparcos errors at epoch (1991.25) updated to our epoch using the Hipparcos 
proper motion errors.  No significant misalignment of the frames was found to
within the formal errors of about 3 mas per axis.  The reduced $\chi^2$ for the 
solutions is $\approx$1.0, confirming the error estimates for the input data. 

\section{PROPER MOTIONS \label{PROP}}

The positions of the 19 radio stars from our VLA+PT observations were 
combined with previous VLA observations \citep{JWFD:85, JDG:03} 
to determine stellar proper motions, $\mu_{\alpha \cos\delta}$ and 
$\mu_{\delta}$, for the 15 sources common to both programs.  
Although the data cover a long time range, 1978--2000, the sampling is not 
sufficient to enable the determination of source parallaxes.  We therefore 
computed proper motions for the 15 stars using the parallaxes obtained from 
Hipparcos.  The proper motions derived from the combined VLA and VLA+PT data 
are listed in Table \ref{PROP_MOT_TAB}.  The values listed in columns 3 and 4 of
Table \ref{PROP_MOT_TAB} were computed using a linear least-squares fit to 
the data weighted by the position errors for each observation.  Position 
errors for the previous VLA observations were estimated to be 30~mas in both 
$\alpha \cos\delta$ and $\delta$ \citep{JDG:03} and we have adopted these values 
in our proper motion analysis.  We did not attempt to 
include accelerations or to model possible companions
in wide orbits with the exception of the star UX Ari discussed in the next 
section.  

In addition to our VLA/VLA+PT proper motions, Table~\ref{PROP_MOT_TAB} 
lists the proper motions derived from the Hipparcos mission and 
from long-term VLBI observations by \cite{LPJPRTRG:99}. Comparing the 
various proper motions listed in the table, one can see that 
the VLA/VLA+PT values, with mean errors of 1.44~mas~yr$^{-1}$ 
in $\mu_{\alpha \cos\delta}$ and 1.79~mas~yr$^{-1}$ in $\mu_{\delta}$,
are beginning to approach the accuracies of the Hipparcos proper motions
with mean errors of 0.95~mas~yr$^{-1}$ and 0.87~mas~yr$^{-1}$ respectively.  
For a few of the stars listed in Table~\ref{PROP_MOT_TAB}, our proper 
motion errors are actually smaller than those derived from the Hipparcos 
observations.  

Comparisons of the VLA/VLA+PT proper motions with those derived from Hipparcos
and VLBI data are shown in Figures~\ref{PROP_MOT_VLA} and \ref{PROP_MOT_VLBI}.
Figure~\ref{PROP_MOT_VLA} plots the differences between the proper motions 
in $\alpha \cos\delta$ ($\Delta\mu_{\alpha \cos\delta}$) and $\delta$ 
($\Delta\mu_{\delta}$), as derived from our VLA/VLA+PT observations 
and those of Hipparcos.  The error bars are the root-sum-square of the uncertainties 
reported for the two sets of proper motions.  For six of the 15 stars common 
to both data sets, proper motions in both $\alpha \cos\delta$ and $\delta$
are in agreement to within the error bars.  An additional three stars agree 
in $\mu_{\alpha \cos\delta}$ only, four stars agree in $\mu_{\delta}$
only, and two stars do not agree at the 1$\sigma$ level.
Figure~\ref{PROP_MOT_VLBI} shows a similar comparison of the VLA/VLA+PT proper 
motions with those derived from VLBI observations of \cite{LPJPRTRG:99}.  
Although the errors reported for the VLBI proper motions are significantly smaller
than those estimated for the VLA/VLA+PT data (see Table~\ref{PROP_MOT_TAB}), the 
computed proper motions are in complete agreement for four of the six stars 
common to both experiments.  The other two stars do not agree within the 
uncertainties in either $\mu_{\alpha \cos\delta}$ or $\mu_{\delta}$.  One 
of these stars is Algol a known ternary, for which we only have 6 data points.
The other star is UX Ari which was mentioned previously as having large 
differences between our VLA+PT position and the Hipparcos position.
UX Ari is discussed in more detail below.  

Finally, in columns 9 and 10 of Table~\ref{PROP_MOT_TAB} we combined the 
VLA/VLA+PT, Hipparcos, and VLBI values by computing weighted mean 
proper motions and associated errors for each of the 15 stars having 
VLA/VLA+PT data.  The combined position was weighted by the errors in the 
individual proper motions derived from each of the three data sets:  
VLA/VLA+PT, Hipparcos, and VLBI (when available).  The error in the 
combined position is just the WRMS of the available VLA/VLA+PT, Hipparcos, 
and/or VLBI errors.  Because of the relatively small errors in the VLBI 
derived proper motions, some of the combined proper motions are heavily 
weighted toward the VLBI value (see $\sigma^2$ CrB for example).  
The accuracies of the averaged proper motions exceed those of the 
VLA/VLA+PT and Hipparcos data alone with average WRMS errors of 
0.47~mas~yr$^{-1}$ in $\mu_{\alpha \cos\delta}$ and 0.48~mas~yr$^{-1}$ 
in $\mu_{\delta}$.

\section{MOTION OF UX ARIETIS \label{DIS_UXARI}}

UX Ari is an active RS CVn binary with an orbital period of 6.44 days
and an inclination of $\approx$60$^{\circ}$.  
Despite being the most observed source in our study, the star UX Ari 
exhibited unexpectedly large position offsets between our VLA+PT measurements 
and those derived from the Hipparcos mission.  
We scrutinized the reduction and analysis of the VLA+PT data for UX Ari,
and could find no errors in the processing.  The positions determined for UX Ari 
when phase-referenced independently to the primary and secondary ICRF calibrators 
agreed to within 10~mas in $\alpha \cos\delta$ and 1~mas in $\delta$.  
In addition, the VLA+PT position of the secondary calibrator phase-referenced 
to the primary calibrator agreed with the "true" ICRF position to within 
8~mas in $\alpha \cos\delta$ and 1~mas in $\delta$.  

Figure \ref{UXARI} 
plots the 14 astrometric position measurements derived from our VLA/VLA+PT 
observations in $\alpha \cos\delta$ and $\delta$ from 1978--2000.  Because
a linear least-squares fit provided a poor representation of the data, 
we decided to include acceleration terms in the least-squares fits
to the time series for UX Ari.  These fits are shown as solid curves in 
Figure \ref{UXARI}.  Central epochs of 1985.2922 in right ascension and 
1985.2311 in declination were computed, and the data were fit with a 
second order polynomial using a weighted least-squares method.  
The linear terms in the fits represent the proper motions which are 
reported in the additional row for UX Ari in Table~\ref{PROP_MOT_TAB}.  
The second order terms represent the accelerations obtained from the fit, 
which are $-0.60\pm 0.03$~mas~yr$^{-2}$ in $\alpha \cos\delta$ and 
$-0.31\pm 0.02$~mas~yr$^{-2}$ in $\delta$.

It should not be too surprising that the observed motion for UX Ari
consists of more than just linear proper motion.  \cite{LPJPRTRG:99} 
reported statistically significant accelerations of 
$-0.54\pm 0.07$~mas~yr$^{-2}$ (8$\sigma$) in $\alpha \cos\delta$ and
$-0.29\pm 0.07$~mas~yr$^{-2}$ (4$\sigma$) in $\delta$
for their VLBI observations.  These accelerations are in very good agreement
with the accelerations derived from our VLA/VLA+PT data.  \cite{LPJPRTRG:99}  
suggested that these accelerations might be due to the gravitational influence
of a companion and, if bound, the wider components should have
an orbital period many times the 11-year span of their data.  The RS CVn system 
does, in fact, have a companion, and UX Ari is listed in the Washington 
Double Star catalog \citep{MWHDW:01} with the designation WDS 03266+2843.  
The separation between the RS CVn system and the third star 
has been measured several times using speckle interferometry, most 
recently by \cite{HMMRTTPLF:00} who measured a separation of $0''.256$ at 
an epoch of 1996.8658.  In addition, spectroscopic radial velocity measurements 
of the RS CVn components by \cite{DA:01} show a systematic variation of the 
center of mass velocity with time indicating the influence 
of a third star.  Their preliminary fits to the radial velocity measurements 
yielded periods of approximately 10.7 and 21.5 years for circular and elliptical 
orbits respectively. 

A fit to the very limited data in the WDS (W. I. Hartkopf 2002, private 
communication) yielded an orbital period of $51.1\pm 24.8$~yr, 
a semimajor axis of $0''.34\pm 0''.11$,
and an inclination of $94.4^{\circ}\pm 3.5^{\circ}$ for the third component of 
the system.  This period is significantly longer than those determined from 
the radial velocity measurements and more in line with expectations given 
the accelerations derived for the RS CVn system from the VLBI measurements 
of \cite{LPJPRTRG:99} and the VLA/VLA+PT data here.  The inclination angle 
estimated from the WDS data is significantly tilted with respect to inclination
angle of the RS CVn orbit of $\approx$60$^{\circ}$.  Assuming the period and 
semimajor axis from the WDS orbit, the total mass of the system is determined
to be $\approx$1.9$ M_{\odot}$, which is slightly less than the mass of the 
RS CVn system alone ($m_{1,2} = 2.05 M_{\odot}$) given in \cite{DA:01}.
A second fit to the WDS data assuming an inclination for the outer orbit 
equal to that of the RS CVn system (C. A. Hummel 2002, private communication)
yielded a smaller semimajor axis, a shorter period, and an 
eccentricity close to 1.  The total mass for this orbit is again less 
than the assumed mass of the RS CVn component
of the system.  The total system masses determined from the fits to the
WDS data clearly do not agree with the mass of the RS CVn system, and 
probably reflect limitations in the orbits determined from the sparse
data.  Additional observations would be very helpful in determining 
a better orbit for the third component of UX Ari and the total mass 
of the system. 

\section{DISCUSSION}

We have determined the astrometric positions for 19 radio stars using the 
VLA+PT configuration. The positions presented here, with uncertainties 
on the order of 10 mas or better, represent a factor of three improvement over 
the previous VLA results \citep{JWFD:85, JDG:03}.  Stellar positions from Hipparcos 
are degrading with time due to errors in the Hipparcos proper motions on the order 
of 1~mas~yr$^{-1}$ and due to unmodeled rotations in the frame with respect to the 
extragalactic objects estimated to be 0.25~mas~yr$^{-1}$ per axis.
Taking into account these uncertainties, for many of the stars in our list, 
our VLA+PT positions are better than the corresponding Hipparcos positions at epoch.  
The proper motions derived from our VLA+PT observations combined with 
previous VLA data of \citeauthor{JDG:03} have errors which are on the 
order of, and in some cases are better than, those obtained from Hipparcos. 

To our knowledge we provide here the first critical check on the
Hipparcos frame to ICRF alignment after the initial effort in the mid 1990's.
The formal, predicted error on the frame alignment at our epoch of 2000.94,
thus 9.69 years after the mean Hipparcos epoch of 1991.25, is 2.5 mas.
Our observations indicate insignificant alignment rotations of
$\leq$2.3~mas with a formal error of $\approx$3~mas per axis.  In a pilot investigation, 
172 extragalactic sources were used to compare the ICRF/optical frames 
(Assafin et al. 2003, AJ, in press).  This program yielded similar results with no
significant system rotations found and formal errors on the 3~mas level.  
However, systematic errors in the preliminary, wide-field, optical data 
of $\approx$10~mas were reported.  For determining possible radio/optical frame 
differences the use of radio stars is currently more competitive because
the accurate Hipparcos data can be utilized directly.

Although future astrometric satellite missions will likely observe distant 
extragalactic objects directly to provide a tie to the radio reference frame,
astrometric observations of radio stars will still provide an important 
verification of such a link as demonstrated here.  It is possible that on the proposed 
microarcsecond scales measurable by future astrometric satellites, the extragalactic sources 
may have significant time dependent source structure affecting quasar optical positions.  
This is indeed the case on milliarcsecond levels at radio wavelengths as shown by
\cite{FC:97} and \cite{FC:00}.  

The VLA+PT configuration provides a useful tool for radio-star astrometry 
because of its ability to do both high resolution and high sensitivity observations 
at the same time.  Compared with the 24 hour VLA+PT observing time, the same 
19 stars plus phase calibrator sources would have taken on the order of 120 
hours of time on the VLBA.  To observe the 50
stars in the \citeauthor{JDG:03} list at multiple epochs using the VLBA would
require a tremendous allocation of time, and such a project is probably impractical.
Clearly additional observations are necessary to update the positions and proper 
motions of the 50 radio stars previously observed with the VLA especially 
since the VLA+PT configuration offers a factor of three improvement over the 
previous measurements.  Additional observations, with better sampling in time, 
will also enable the determination of the parallaxes directly from the radio data.

There are several areas in which we might extend or improve our radio-star 
astrometry in future experiments.  As mentioned in section 3, our use 
of the dirty rather than CLEANed images to determine the positions may have been 
unnecessarily conservative, since there was little difference between the 
positions derived from the two sets of images.  In Figure~\ref{ELEVERROR} there
was a slight increase in position error as a function of decreasing source 
elevation likely due to unmodeled atmospheric and ionospheric effects.  Modeling of the 
phase fluctuations due to these phenomena may result in improved astrometric
positions.  Observationally, the use of fast-switching and two phase calibrators per 
target star proved to be beneficial in our analysis of the data and worth
the additional observation time.  In future experiments, it may be useful to increase 
the number of observations per star to a number greater than the three we used 
for this experiment.  For example, \cite{JDG:03} used 5--6 observations per star
in their VLA radio-star astrometry.  The three observations per star used here was 
the result of a compromise between our desire for better $uv$ coverage and the scheduling 
constraints imposed by the use of fast-switching and the VLA+PT configuration.  
Finally, the greatest improvements in stellar astrometry at radio wavelengths will 
likely come from future enhancements to the VLA.  The increased sensitivity 
of Phase I of the EVLA project will extend the total number of stars 
observable with the VLA and will shorten time required to observe these stars.  
Phase II of the EVLA project will further increase the resolution provided by 
the VLA+PT configuration with the addition of up to 8 new antennas at distances 
of up to 300 km from the VLA.

\begin{acknowledgements}
The authors would like the thank Dr. William Hartkopf and Dr. Christian Hummel
for providing orbital fits to the WDS data for the star UX Ari.
\end{acknowledgements}

\clearpage

\centering{Figure Captions}

\figcaption{Distribution of the 19 observed radio stars plotted on an Aitoff equal-area
projection of the celestial sphere.  The dotted line represents the Galactic equator. 
\label{AITOFF}}

\figcaption{Formal errors from the 2-D Gaussian fits to the dirty images for each 
observation plotted as a function of source elevation.  Errors in $\alpha \cos\delta$ 
are plotted in (a) and errors in declination are plotted in (b).\label{ELEVERROR}}

\figcaption{Differences between the ICRF reference positions and our VLA+PT measured
positions as a function of source right ascension $\alpha$ for the 38 observed quasars.
Differences in $\alpha \cos\delta$ are plotted in (a) and differences in declination
$\delta$ are plotted in (b). Error bars are from our VLA+PT measurements.
\label{ICRFVLARA}}

\figcaption{Differences between the ICRF reference positions and our VLA+PT measured
positions as a function of source declination $\delta$ for the 38 observed quasars.
Differences in $\alpha \cos\delta$ are plotted in (a) and differences in declination
are plotted in (b). Error bars are from our VLA+PT measurements. 
\label{ICRFVLADEC}}

\figcaption{Differences between the Hipparcos positions updated to the epoch of our 
observations, and our VLA+PT measured positions as a function of source right 
ascension $\alpha$ for the 19 radio stars observed.  Differences in $\alpha \cos\delta$ 
are plotted in (a) and differences in declination are plotted in (b). Error bars are 
from our VLA+PT measurements. \label{HIPVLARA}}

\figcaption{Differences between the Hipparcos positions updated to the epoch of our 
observations, and our VLA+PT measured positions as a function of source declination
$\delta$ for the 19 radio stars observed.  Differences in $\alpha \cos\delta$ 
are plotted in (a) and differences in declination are plotted in (b). Error bars are 
from our VLA+PT measurements. \label{HIPVLADEC}}

\figcaption{Differences in the proper motions, $\Delta\mu_{\alpha \cos\delta}$ vs. 
$\Delta\mu_{\delta}$, as derived from our VLA+PT observations and from the Hipparcos
mission.  Error bars are the root-sum-square of the errors give in Table 
\ref{PROP_MOT_TAB} for the two measurement sets. \label{PROP_MOT_VLA}}

\figcaption{Differences in the proper motions, $\Delta\mu_{\alpha \cos\delta}$ vs. 
$\Delta\mu_{\delta}$, as derived from our VLA+PT observations and the VLBI observations
of \cite{LPJPRTRG:99}.  Error bars are the root-sum-square of the errors give in Table 
\ref{PROP_MOT_TAB} for the two measurement sets. \label{PROP_MOT_VLBI}}

\figcaption{Proper motions for the star UX Ari in (a) right ascension and 
(b) declination estimated from our VLA+PT observations and the VLA data of \cite{JDG:03}.  
The solid curve represents a weighted least-squares fit of a second order polynomial
to the data.  Errors for the \citeauthor{JDG:03} VLA data are 30~mas in both 
$\alpha \cos\delta$ and $\delta$. 
\label{UXARI}}

\clearpage

\begin{deluxetable}{lcccrrc}
\tabletypesize{\footnotesize}
\tablewidth{0pt}
\tablecaption{Radio stars and their corresponding ICRF reference sources.\label{SOURCES}}
\tablehead{ Star & \colhead{Hipparcos} & \colhead{Reference} &  \colhead{ICRF}
& \colhead{$\alpha$ (J2000)\tablenotemark{c}} & \colhead{$\delta$ (J2000)\tablenotemark{c}} 
& \colhead{Separation} \\
    & \colhead{Number} & \colhead{Source \tablenotemark{a}} 
& \colhead{Designation\tablenotemark{b}} & \colhead{(h m s)} 
& \colhead{($^{\circ}$ $'$ $''$)} & \colhead{($^{\circ}$)} }
\startdata
LSI61303       &   12469  &   0302+625 &  C  &  03 06 42.659558  &  62 43 02.02417  &   3.5 \\
               &          &   0241+622 &  C  &  02 44 57.696827  &  62 28 06.51459  &   1.3 \\
Algol          &   14576  &   0309+411 &  D  &  03 13 01.962129  &  41 20 01.18353  &   1.0 \\
               &          &   0248+430 &  D  &  02 51 34.536779  &  43 15 15.82858  &   3.9 \\
UX Ari         &   16042  &   0333+321 &  O  &  03 36 30.107599  &  32 18 29.34239  &   4.2 \\
               &          &   0326+277 &  O  &  03 29 57.669413  &  27 56 15.49914  &   1.1 \\
HR1099         &   16846  &   0336-019 &  C  &  03 39 30.937787  & -01 46 35.80399  &   2.5 \\
               &          &   0305+039 &  N  &  03 08 26.223804  &  04 06 39.30105  &   7.9 \\
B Per          &   20070  &   0420+417 &  C  &  04 23 56.009795  &  41 50 02.71277  &   8.5 \\
               &          &   0300+470 &  O  &  03 03 35.242226  &  47 16 16.27546  &  12.3 \\
T Tau          &   20390  &   0409+229 &  N  &  04 12 43.666851  &  23 05 05.45299  &   4.2 \\
               &          &   0400+258 &  D  &  04 03 05.586048  &  26 00 01.50274  &   7.9 \\
$\alpha$ Ori   &   27989  &   0611+131 &  C  &  06 13 57.692766  &  13 06 45.40116  &   7.4 \\
               &          &   0529+075 &  C  &  05 32 38.998531  &  07 32 43.34586  &   5.6 \\
HD50896-N      &   33165  &   0646-306 &  C  &  06 48 14.096441  & -30 44 19.65940  &   6.9 \\
               &          &   0727-115 &  O  &  07 30 19.112472  & -11 41 12.60048  &  14.8 \\
KQ Pup         &   36773  &   0727-115 &  O  &  07 30 19.112472  & -11 41 12.60048  &   3.0 \\
               &          &   0733-174 &  D  &  07 35 45.812508  & -17 35 48.50131  &   3.1 \\
54 Cam         &   39348  &   0749+540 &  D  &  07 53 01.384573  &  53 52 59.63716  &   3.6 \\
               &          &   0831+557 &  D  &  08 34 54.903997  &  55 34 21.07080  &   4.7 \\
TY Pix-N       &   44164  &   0925-203 &  C  &  09 27 51.824323  & -20 34 51.23266  &   9.5 \\
               &          &   0919-260 &  O  &  09 21 29.353874  & -26 18 43.38604  &   5.0 \\
RS CVn         &   64293  &   1308+326 &  D  &  13 10 28.663845  &  32 20 43.78295  &   3.6 \\
               &          &   1404+286 &  O  &  14 07 00.394410  &  28 27 14.68998  &  13.6 \\
HR5110         &   66257  &   1315+346 &  C  &  13 17 36.494189  &  34 25 15.93266  &   4.4 \\
               &          &   1404+286 &  O  &  14 07 00.394410  &  28 27 14.68998  &  10.8 \\
$\delta$ Lib   &   73473  &   1511-100 &  C  &  15 13 44.893444  & -10 12 00.26435  &   3.6 \\
               &          &   1510-089 &  O  &  15 12 50.532939  & -09 05 59.82950  &   3.0 \\
$\sigma^2$ CrB &   79607  &   1611+343 &  C  &  16 13 41.064249  &  34 12 47.90909  &   0.4 \\
               &          &   1600+335 &  D  &  16 02 07.263468  &  33 26 53.07267  &   2.6 \\
$\beta$ Lyra   &   92420  &   1901+319 &  O  &  19 02 55.938870  &  31 59 41.70209  &   3.0 \\
               &          &   1751+288 &  O  &  17 53 42.473634  &  28 48 04.93908  &  12.6 \\
HD199178       &  103144  &   2037+511 &  D  &  20 38 37.034755  &  51 19 12.66269  &   7.5 \\
               &          &   2100+468 &  C  &  21 02 17.056042  &  47 02 16.25451  &   3.0 \\
AR Lac         &  109303  &   2200+420 &  O  &  22 02 43.291377  &  42 16 39.97994  &   3.6 \\
               &          &   2214+350 &  C  &  22 16 20.009910  &  35 18 14.18056  &  10.5 \\
IM Peg         &  112997  &   2250+190 &  N  &  22 53 07.369176  &  19 42 34.62843  &   2.9 \\
               &          &   2251+158 &  O  &  22 53 57.747932  &  16 08 53.56089  &   0.7 \\
\enddata
\tablenotetext{a}{For each star the primary (fast-switched) reference source is listed first 
followed by the secondary reference source.}
\tablenotetext{b}{ICRF source designation \citep{IERS:99}: D = defining, C = candidate, O = other, N = new
in ICRF Extension 1.}
\tablenotetext{c}{ICRF Extension 1 source positions \citep{IERS:99}.}\end{deluxetable}

\clearpage 

\begin{deluxetable}{lccrr}
\tabletypesize{\footnotesize}
\tablewidth{0pt}
\tablecaption{Source positions estimated from the VLA+PT data\label{POSITIONS}}
\tablehead{ Source \tablenotemark{a} & \colhead{$\alpha$ (J2000)} & \colhead{$\delta$ (J2000)} 
& \colhead{S}  &  S.N.R. \\
     & \colhead{(h m s)} & \colhead{($^{\circ}$ $'$ $''$)} & \colhead{(mJy)} &  }
\startdata
LSI61303 & 02 40 31.6646 $\pm$0.0008 ($\pm$0.006$''$) & 61 13 45.593 $\pm$0.003  &  42.2  & 227.8 \\
0302+625 & 03 06 42.6598 $\pm$0.0004 ($\pm$0.003$''$) & 62 43 02.026 $\pm$0.004  & 242.6  & 123.8 \\
0241+621 & 02 44 57.6972 $\pm$0.0021 ($\pm$0.014$''$) & 62 28 06.512 $\pm$0.008  & 605.9  & 142.6 \\
         &                                            &                          &        &       \\ 
Algol    & 03 08 10.1307 $\pm$0.0001 ($\pm$0.001$''$) & 40 57 20.345 $\pm$0.007  &  24.2  & 101.5 \\
0309+411 & 03 13 01.9622 $\pm$0.0013 ($\pm$0.014$''$) & 41 20 01.186 $\pm$0.004  & 454.5  & 132.5 \\
0248+430 & 02 51 34.5361 $\pm$0.0010 ($\pm$0.011$''$) & 43 15 15.824 $\pm$0.006  & 893.8  & 113.6 \\
         &                                            &                          &        &       \\ 
UX Ari   & 03 26 35.3849 $\pm$0.0006 ($\pm$0.008$''$) & 28 42 54.176 $\pm$0.004  &  10.1  & 121.7 \\
0333+321 & 03 36 30.1063 $\pm$0.0006 ($\pm$0.008$''$) & 32 18 29.347 $\pm$0.005  &1515.4  & 153.6 \\
0326+277 & 03 29 57.6700 $\pm$0.0008 ($\pm$0.011$''$) & 27 56 15.500 $\pm$0.005  & 613.1  & 121.4 \\
         &                                            &                          &        &       \\ 
HR1099   & 03 36 47.2869 $\pm$0.0003 ($\pm$0.005$''$) & 00 35 15.772 $\pm$0.005  &  13.5  & 204.1 \\
0336-019 & 03 39 30.9380 $\pm$0.0014 ($\pm$0.020$''$) &-01 46 35.785 $\pm$0.034  &1798.1  &  99.8 \\
0305+039 & 03 08 26.2240 $\pm$0.0011 ($\pm$0.016$''$) & 04 06 39.288 $\pm$0.026  & 562.6  &  84.7 \\
         &                                            &                          &        &       \\ 
B Per    & 04 18 14.6216 $\pm$0.0014 ($\pm$0.014$''$) & 50 17 43.766 $\pm$0.021  &   0.4  &  10.5 \\
0420+417 & 04 23 56.0106 $\pm$0.0008 ($\pm$0.009$''$) & 41 50 02.720 $\pm$0.007  &1040.2  &  61.0 \\
0300+470 & 03 03 35.2409 $\pm$0.0015 ($\pm$0.015$''$) & 47 16 16.259 $\pm$0.008  & 987.6  &  69.6 \\
         &                                            &                          &        &       \\ 
T Tau-N  & 04 21 59.4345 $\pm$0.0004 ($\pm$0.005$''$) & 19 32 06.406 $\pm$0.009  &   1.1  &  21.6 \\
T Tau-S  & 04 21 59.4258 $\pm$0.0005 ($\pm$0.007$''$) & 19 32 05.718 $\pm$0.008  &   6.1  & 131.4 \\
0409+229 & 04 12 43.6669 $\pm$0.0005 ($\pm$0.007$''$) & 23 05 05.445 $\pm$0.009  & 326.5  & 114.4 \\
0400+258 & 04 03 05.5854 $\pm$0.0008 ($\pm$0.010$''$) & 26 00 01.503 $\pm$0.003  & 884.0  &  69.6 \\
         &                                            &                          &        &       \\ 
$\alpha$ Ori & 05 55 10.3061 $\pm$0.0009 ($\pm$0.013$''$) & 07 24 25.432 $\pm$0.026  &   2.8  &  50.2 \\
0611+131 & 06 13 57.6933 $\pm$0.0020 ($\pm$0.029$''$) & 13 06 45.401 $\pm$0.006  & 266.2  &  48.6 \\
0529+075 & 05 32 39.0000 $\pm$0.0017 ($\pm$0.025$''$) & 07 32 43.336 $\pm$0.024  & 903.4  &  56.2 \\
         &                                            &                          &        &       \\ 
HD50896-N& 06 54 13.0456 $\pm$0.0031 ($\pm$0.043$''$) &-23 55 41.993 $\pm$0.107  &   1.1  &  22.3 \\
0646-306 & 06 48 14.1014 $\pm$0.0017 ($\pm$0.022$''$) &-30 44 19.634 $\pm$0.027  & 527.7  &  28.5 \\
0727-115 & 07 30 19.1085 $\pm$0.0018 ($\pm$0.026$''$) &-11 41 12.589 $\pm$0.100  &2026.6  &  24.0 \\
         &                                            &                          &        &       \\ 
KQ Pup   & 07 33 47.9637 $\pm$0.0002 ($\pm$0.002$''$) &-14 31 25.994 $\pm$0.007  &   2.3  &  41.7 \\
0727-115 & 07 30 19.1124 $\pm$0.0016 ($\pm$0.023$''$) &-11 41 12.607 $\pm$0.008  &3869.7  &  44.8 \\
0733-174 & 07 35 45.8133 $\pm$0.0006 ($\pm$0.008$''$) &-17 35 48.483 $\pm$0.010  &1002.5  &  49.7 \\
         &                                            &                          &        &       \\ 
54 Cam   & 08 02 35.7815 $\pm$0.0012 ($\pm$0.010$''$) & 57 16 24.997 $\pm$0.005  &   1.1  &  26.4 \\
0749+540 & 07 53 01.3851 $\pm$0.0014 ($\pm$0.012$''$) & 53 52 59.632 $\pm$0.029  & 854.6  & 105.0 \\
0831+557 & 08 34 54.9023 $\pm$0.0014 ($\pm$0.012$''$) & 55 34 21.079 $\pm$0.014  &2641.6  & 137.9 \\
         &                                            &                          &        &       \\ 
TY Pix-N & 08 59 42.7205 $\pm$0.0003 ($\pm$0.005$''$) &-27 48 58.711 $\pm$0.014  &   2.1  &  40.8 \\
0925-203 & 09 27 51.8231 $\pm$0.0011 ($\pm$0.015$''$) &-20 34 51.246 $\pm$0.033  & 325.5  &  78.4 \\
0919-260 & 09 21 29.3552 $\pm$0.0007 ($\pm$0.010$''$) &-26 18 43.370 $\pm$0.024  &1590.4  &  72.4 \\
         &                                            &                          &        &       \\ 
RS CVn   & 13 10 36.9034 $\pm$0.0014 ($\pm$0.017$''$) & 35 56 05.604 $\pm$0.004  &   1.2  &  32.3 \\
1308+326 & 13 10 28.6638 $\pm$0.0007 ($\pm$0.009$''$) & 32 20 43.803 $\pm$0.032  &1397.0  &  89.0 \\
1404+286 & 14 07 00.3940 $\pm$0.0017 ($\pm$0.022$''$) & 28 27 14.679 $\pm$0.030  &1815.0  &  99.7 \\
         &                                            &                          &        &       \\ 
HR5110   & 13 34 47.8155 $\pm$0.0003 ($\pm$0.004$''$) & 37 10 56.672 $\pm$0.010  &   7.2  & 116.7 \\
1315+346 & 13 17 36.4934 $\pm$0.0020 ($\pm$0.025$''$) & 34 25 15.946 $\pm$0.039  & 239.0  &  82.4 \\
1404+286 & 14 07 00.3942 $\pm$0.0034 ($\pm$0.045$''$) & 28 27 14.672 $\pm$0.040  &1715.0  &  78.0 \\
         &                                            &                          &        &       \\ 
$\delta$ Lib & 15 00 58.3456 $\pm$0.0009 ($\pm$0.013$''$) &-08 31 08.219 $\pm$0.018  &   1.4  &  37.5 \\
1511-100 & 15 13 44.8943 $\pm$0.0012 ($\pm$0.018$''$) &-10 12 00.255 $\pm$0.007  & 813.0  & 100.4 \\
1510-089 & 15 12 50.5324 $\pm$0.0005 ($\pm$0.008$''$) &-09 05 59.837 $\pm$0.005  &1010.0  & 120.2 \\
         &                                            &                          &        &       \\ 
$\sigma^2$ CrB & 16 14 40.8337 $\pm$0.0001 ($\pm$0.002$''$) & 33 51 30.892 $\pm$0.002  &  12.9  &  68.0 \\
1611+343 & 16 13 41.0636 $\pm$0.0007 ($\pm$0.009$''$) & 34 12 47.910 $\pm$0.005  &4295.0  & 163.9 \\
1600+335 & 16 02 07.2638 $\pm$0.0006 ($\pm$0.007$''$) & 33 26 53.073 $\pm$0.006  &1271.0  & 198.6 \\
         &                                            &                          &        &       \\ 
$\beta$ Lyra & 18 50 04.7945 $\pm$0.0006 ($\pm$0.007$''$) & 33 21 45.595 $\pm$0.004  &   4.3  & 101.7 \\
1901+319 & 19 02 55.9403 $\pm$0.0021 ($\pm$0.027$''$) & 31 59 41.718 $\pm$0.014  &1005.0  & 102.6 \\
1751+288 & 17 53 42.4723 $\pm$0.0016 ($\pm$0.022$''$) & 28 48 04.925 $\pm$0.015  & 664.0  & 100.6 \\
         &                                            &                          &        &       \\ 
HD199178 & 20 53 53.6544 $\pm$0.0006 ($\pm$0.007$''$) & 44 23 11.080 $\pm$0.003  &   5.0  & 120.1 \\
2037+511 & 20 38 37.0331 $\pm$0.0017 ($\pm$0.016$''$) & 51 19 12.655 $\pm$0.008  &2503.7  & 160.5 \\
2100+468 & 21 02 17.0567 $\pm$0.0014 ($\pm$0.014$''$) & 47 02 16.265 $\pm$0.012  & 128.8  &  15.6 \\
         &                                            &                          &        &       \\ 
AR Lac   & 22 08 40.8114 $\pm$0.0002 ($\pm$0.002$''$) & 45 44 32.147 $\pm$0.001  &   3.4  & 147.1 \\
2200+420 & 22 02 43.2910 $\pm$0.0008 ($\pm$0.009$''$) & 42 16 39.974 $\pm$0.004  &2501.0  & 188.0 \\
2214+350 & 22 16 20.0102 $\pm$0.0006 ($\pm$0.008$''$) & 35 18 14.184 $\pm$0.005  & 465.1  & 164.9 \\
         &                                            &                          &        &       \\ 
IM Peg   & 22 53 02.2638 $\pm$0.0003 ($\pm$0.004$''$) & 16 50 28.271 $\pm$0.016  &   0.4  & 120.9 \\
2250+190 & 22 53 07.3692 $\pm$0.0006 ($\pm$0.008$''$) & 19 42 34.628 $\pm$0.003  & 414.6  & 158.9 \\
2251+158 & 22 53 57.7480 $\pm$0.0005 ($\pm$0.007$''$) & 16 08 53.559 $\pm$0.009  &9109.8  & 198.9 \\
\enddata
\tablenotetext{a}{For the three stars HD50896-N, RS CVn, and HR5110, the star's position was determined 
using the secondary (not fast-switched) ICRF calibrator as the phase reference.}
\end{deluxetable}

\clearpage

\begin{deluxetable}{lcc}
\tabletypesize{\scriptsize}
\tablewidth{0pt}
\tablecaption{Radio/Optical Frame Alignment. \label{ROT_TAB}}
\tablehead{\colhead{    } &  \colhead{Rotation} & \colhead{Formal Error} \\
           \colhead{Axis} &  \colhead{(mas)}    & \colhead{(mas)}} 
\startdata
x \dotfill &  $-$0.2 &  2.9  \\
y \dotfill &  $-$1.9 &  3.2  \\
z \dotfill &     2.3 &  2.8  \\
\enddata
\tablenotetext{a}{Mean error of unit weight for the alignment is 9.8~mas, with a 
reduced $\chi^2$ equal to 0.98.}
\end{deluxetable}

\clearpage

\begin{deluxetable}{lcrrrrrrrr}
\tabletypesize{\footnotesize}
\tablewidth{0pt}
\rotate
\tablecaption{Comparison of radio-star proper motions. \label{PROP_MOT_TAB}}
\tablehead{
\colhead{}     & \colhead{Number} & \multicolumn{2}{c}{VLA/VLA+PT Proper Motions \tablenotemark{b}} & 
\multicolumn{2}{c}{Hipparcos Proper Motions} & \multicolumn{2}{c}{VLBI Proper Motions \tablenotemark{c}}  &
\multicolumn{2}{c}{Combined Proper Motions \tablenotemark{d}} \\
{Star} & \colhead{of} & \colhead{$\mu_{\alpha \cos\delta}$} & \colhead{$\mu_{\delta}$} & \colhead{$\mu_{\alpha \cos\delta}$} & 
\colhead{$\mu_{\delta}$} & \colhead{$\mu_{\alpha \cos\delta}$} & \colhead{$\mu_{\delta}$} &
\colhead{$\mu_{\alpha \cos\delta}$} & \colhead{$\mu_{\delta}$}\\
\colhead{}     & \colhead{Meas. \tablenotemark{a}} & \colhead{(mas~yr$^{-1}$)} 
& \colhead{(mas~yr$^{-1}$)} & \colhead{(mas~yr$^{-1}$)} & 
\colhead{(mas~yr$^{-1}$)} & \colhead{(mas~yr$^{-1}$)} & \colhead{(mas~yr$^{-1}$)} & 
\colhead{(mas~yr$^{-1}$)} & \colhead{(mas~yr$^{-1}$)}
}
\startdata
LSI61303       &   1   &   \nodata          &  \nodata           &    0.62 $\pm$ 1.95 &    1.63 $\pm$ 1.75 
&    0.97 $\pm$ 0.26 &   -1.21 $\pm$ 0.32   &     \nodata   &   \nodata \\

Algol          &   6   &    3.67 $\pm$ 0.55 &   -3.26 $\pm$ 0.87 &    2.39 $\pm$ 0.77 &   -1.44 $\pm$ 0.88 
&    2.79 $\pm$ 0.14 &   -0.64 $\pm$ 0.18   &      2.83 $\pm$ 0.13   &   -0.77 $\pm$ 0.17\\

UX Ari         &   14  &   38.83 $\pm$ 0.54 & -105.48 $\pm$ 0.54 &   41.35 $\pm$ 1.41 & -104.29 $\pm$ 1.35 
&   41.23 $\pm$ 0.18 & -104.01 $\pm$ 0.20   &     40.98 $\pm$ 0.17  &  -104.19 $\pm$ 0.19\\

UX Ari\tablenotemark{e} &   14  &   45.64 $\pm$ 0.47 & -101.84 $\pm$ 0.25 &  \nodata &  \nodata 
&   \nodata & \nodata   &  \nodata  &  \nodata \\

HR1099         &   13  &  -31.87 $\pm$ 0.41 & -161.09 $\pm$ 0.52 &  -32.98 $\pm$ 0.93 & -163.45 $\pm$ 0.88 
&  -31.59 $\pm$ 0.33 & -161.69 $\pm$ 0.31   &    -31.79 $\pm$ 0.25  &  -161.70 $\pm$ 0.25\\

B Per          &   5   &   44.24 $\pm$ 2.58 &  -57.33 $\pm$ 1.97 &   46.59 $\pm$ 1.17 &  -56.43 $\pm$ 0.94 
&   \nodata          &   \nodata            &     46.19 $\pm$ 1.07  &   -56.60 $\pm$ 0.85\\

T Tau-N        &   1   &   \nodata          &  \nodata           &   15.45 $\pm$ 1.88 &  -12.48 $\pm$ 1.62 
&   \nodata          &   \nodata            &     \nodata   &   \nodata \\

$\alpha$ Ori   &   3   &   24.17 $\pm$ 1.24 &   10.07 $\pm$ 1.81 &   27.33 $\pm$ 2.30 &   10.86 $\pm$ 1.46 
&   \nodata          &   \nodata            &     24.88 $\pm$ 1.09  &    10.55 $\pm$ 1.14\\

HD50896-N      &   2   &   -1.19 $\pm$ 1.86 &    9.32 $\pm$ 6.44 &   -3.86 $\pm$ 0.43 &    4.75 $\pm$ 0.66 
&   \nodata          &   \nodata            &     -3.73 $\pm$ 0.42  &    4.80 $\pm$ 0.66\\

KQ Pup         &   4   &   -8.55 $\pm$ 1.43 &    8.24 $\pm$ 1.67 &   -7.73 $\pm$ 0.64 &    3.62 $\pm$ 0.53 
&   \nodata          &   \nodata            &     -7.87 $\pm$ 0.58  &     4.04 $\pm$ 0.51\\

54 Cam         &   2   &  -35.34 $\pm$ 2.98 &  -57.13 $\pm$ 1.64 &  -38.28 $\pm$ 0.78 &  -59.08 $\pm$ 0.63 
&   \nodata          &   \nodata            &    -38.10 $\pm$ 0.76  &   -58.83 $\pm$ 0.59\\

TY Pix-N       &   2   &  -45.25 $\pm$ 1.32 &  -44.13 $\pm$ 1.92 &  -43.99 $\pm$ 0.47 &  -44.80 $\pm$ 0.55 
&   \nodata          &   \nodata            &    -44.13 $\pm$ 0.44  &   -44.75 $\pm$ 0.53\\

RS CVn         &   4   &  -50.61 $\pm$ 1.32 &   27.05 $\pm$ 1.52 &  -49.14 $\pm$ 0.88 &   21.49 $\pm$ 0.72 
&   \nodata          &   \nodata            &    -49.61 $\pm$ 0.73  &    22.51 $\pm$ 0.65\\

HR5110         &   5   &   84.53 $\pm$ 1.42 &   -9.38 $\pm$ 1.30 &   84.70 $\pm$ 0.45 &   -9.81 $\pm$ 0.39 
&   85.50 $\pm$ 0.13 &   -9.22 $\pm$ 0.16   &     85.43 $\pm$ 0.13  &    -9.31 $\pm$ 0.15\\

$\delta$ Lib   &   4   &  -66.00 $\pm$ 1.84 &   -5.05 $\pm$ 2.35 &  -66.20 $\pm$ 0.86 &   -3.40 $\pm$ 0.81 
&   \nodata          &   \nodata            &    -66.16 $\pm$ 0.78  &    -3.58 $\pm$ 0.77\\

$\sigma^2$ CrB &   2   & -266.66 $\pm$ 1.21 &  -86.69 $\pm$ 1.62 & -266.47 $\pm$ 0.86 &  -86.88 $\pm$ 1.12 
& -267.05 $\pm$ 0.04 &  -86.66 $\pm$ 0.05   &    -267.05 $\pm$ 0.04 &   -86.66 $\pm$ 0.05\\

$\beta$ Lyra   &   4   &    2.79 $\pm$ 1.38 &   -5.24 $\pm$ 1.18 &    1.10 $\pm$ 0.44 &   -4.46 $\pm$ 0.51 
&   \nodata          &   \nodata            &       1.26 $\pm$ 0.42 &    -4.58 $\pm$ 0.47\\

HD199178       &   1   &   \nodata          &   \nodata          &   26.77 $\pm$ 0.77 &   -1.15 $\pm$ 0.61 
&   26.60 $\pm$ 0.41 &   -1.24 $\pm$ 0.43   &      \nodata   &   \nodata \\

AR Lac         &   5   &  -51.21 $\pm$ 1.50 &   47.96 $\pm$ 1.45 &  -52.48 $\pm$ 0.46 &   47.88 $\pm$ 0.53 
&  -52.08 $\pm$ 0.13 &   47.03 $\pm$ 0.19   &     -52.10 $\pm$ 0.12 &    47.14 $\pm$ 0.18\\

IM Peg         &   1   &   \nodata          &   \nodata          &  -20.97 $\pm$ 0.61 &  -27.59 $\pm$ 0.57 
&  -20.59 $\pm$ 0.46 &  -27.53 $\pm$ 0.40   &      \nodata   &   \nodata \\
\enddata
\tablenotetext{a}{Total number of position measurements from our VLA+PT observations plus previous VLA observations
\citep{JWFD:85, JDG:03}.}
\tablenotetext{b}{Proper motions derived from combined VLA and VLA+PT observations.}
\tablenotetext{c}{Proper motions from \cite{LPJPRTRG:99}.}
\tablenotetext{d}{Weighted mean of the proper motions from VLA/VLA+PT, Hipparcos, 
and \cite{LPJPRTRG:99} data.}
\tablenotetext{e}{UX Ari proper motions with acceleration terms included.}
\end{deluxetable}

\clearpage

\begin{figure}[hbt]
\plotone{boboltz.fig1.eps}

\centerline{Figure 1}
\end{figure}

\begin{figure}[hbt]
\plotone{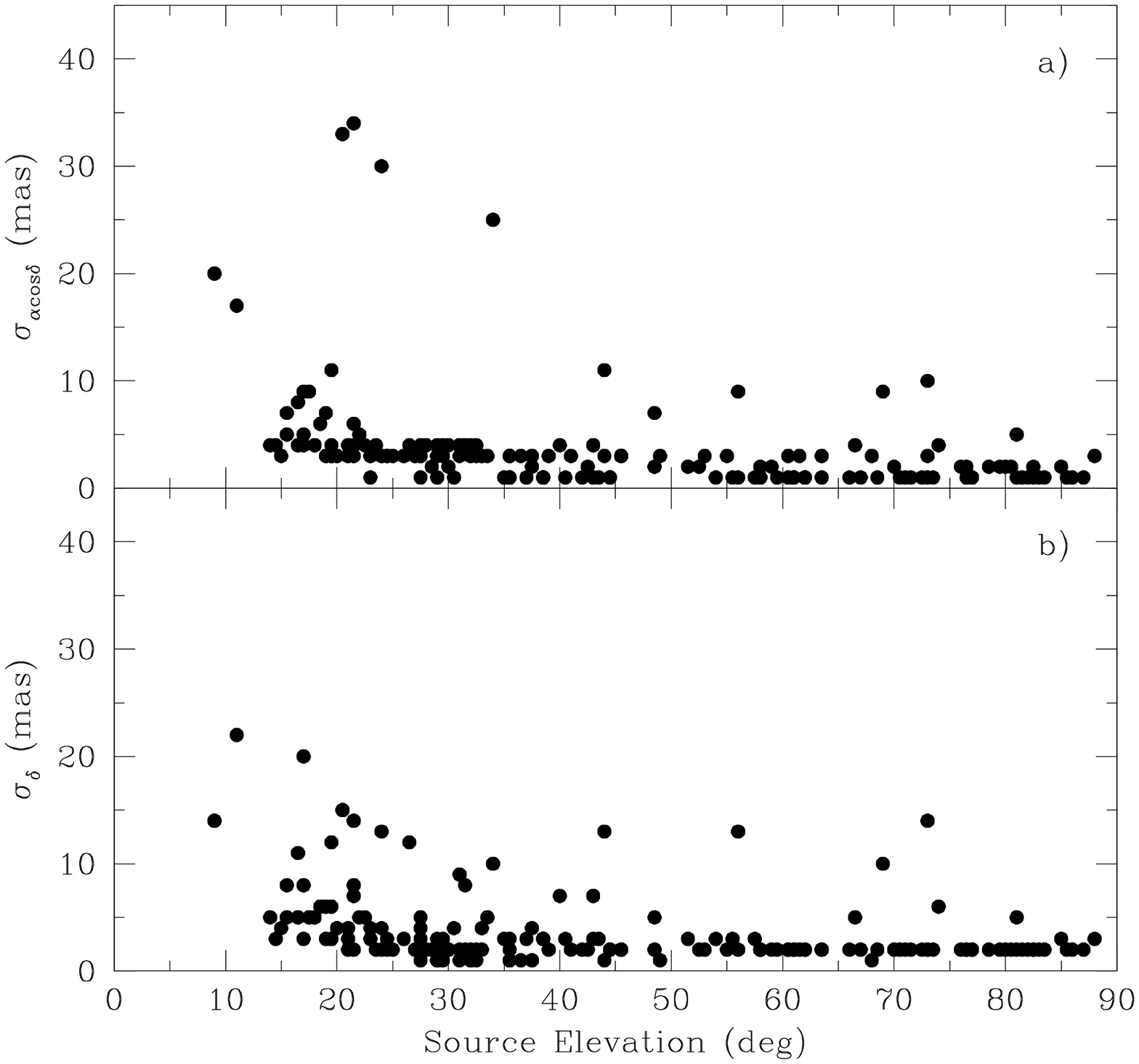}

\centerline{Figure 2}
\end{figure}

\begin{figure}[hbt]
\plotone{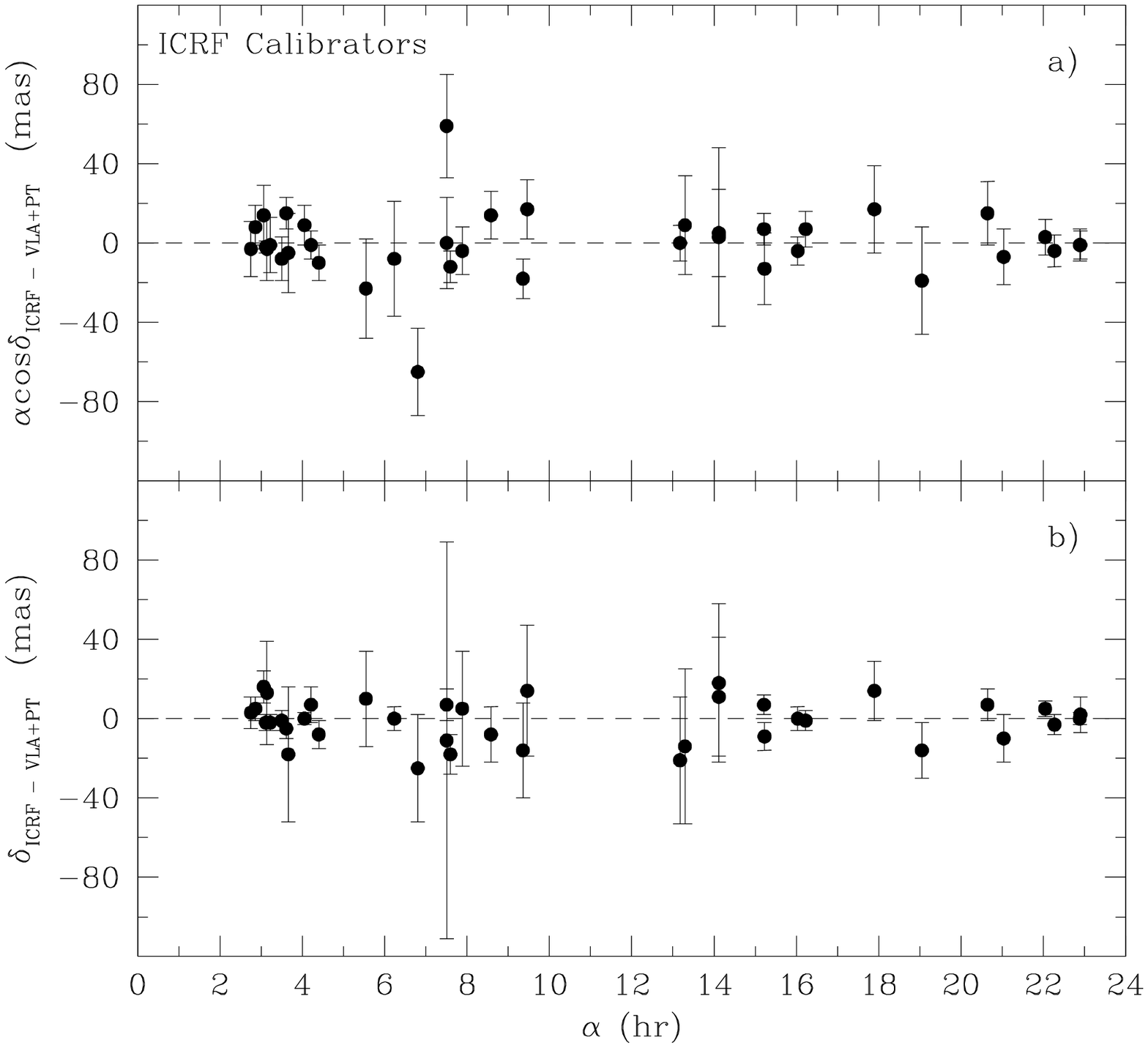}

\centerline{Figure 3}
\end{figure}

\begin{figure}[hbt]
\plotone{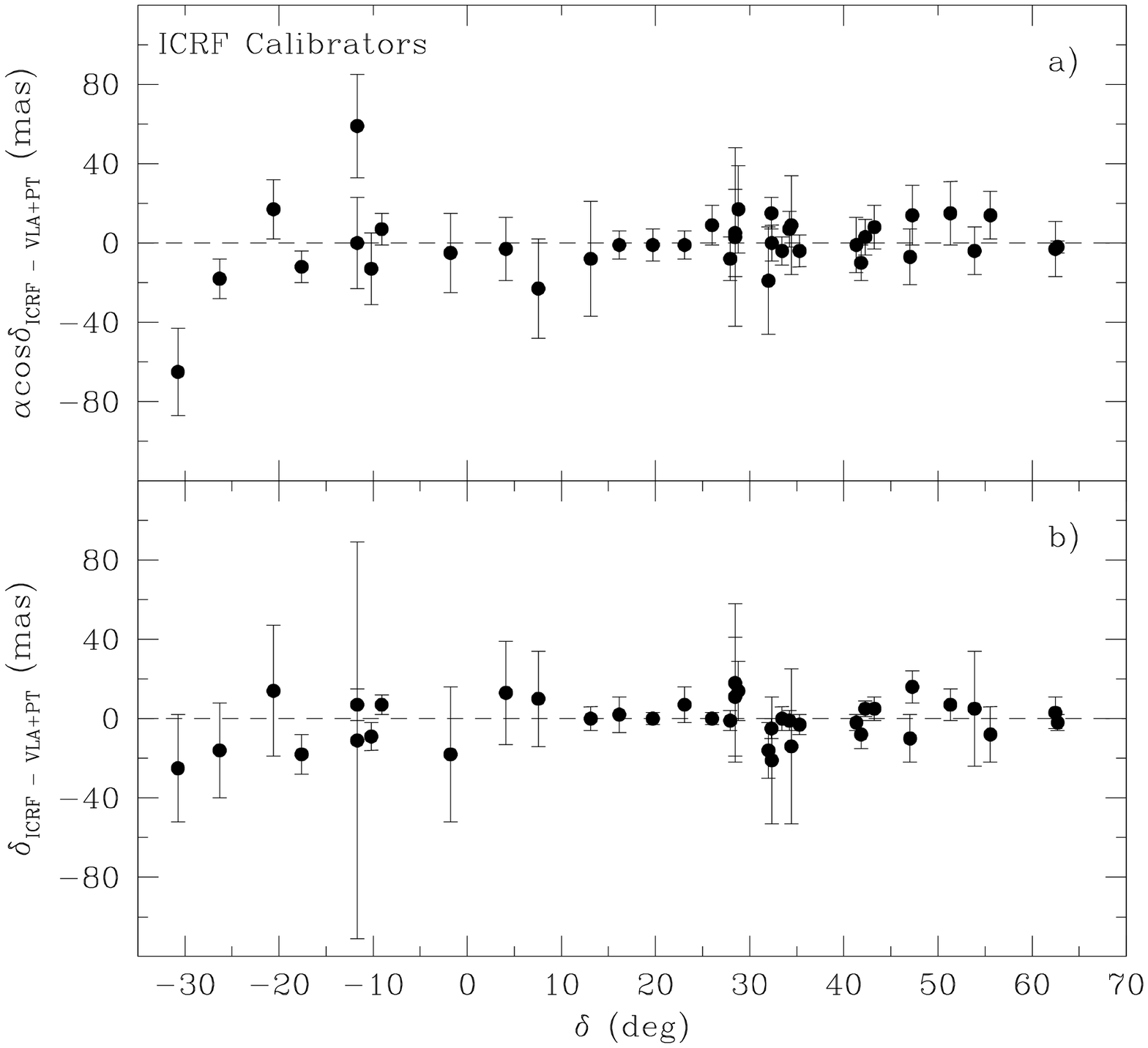}

\centerline{Figure 4}
\end{figure}

\begin{figure}[hbt]
\plotone{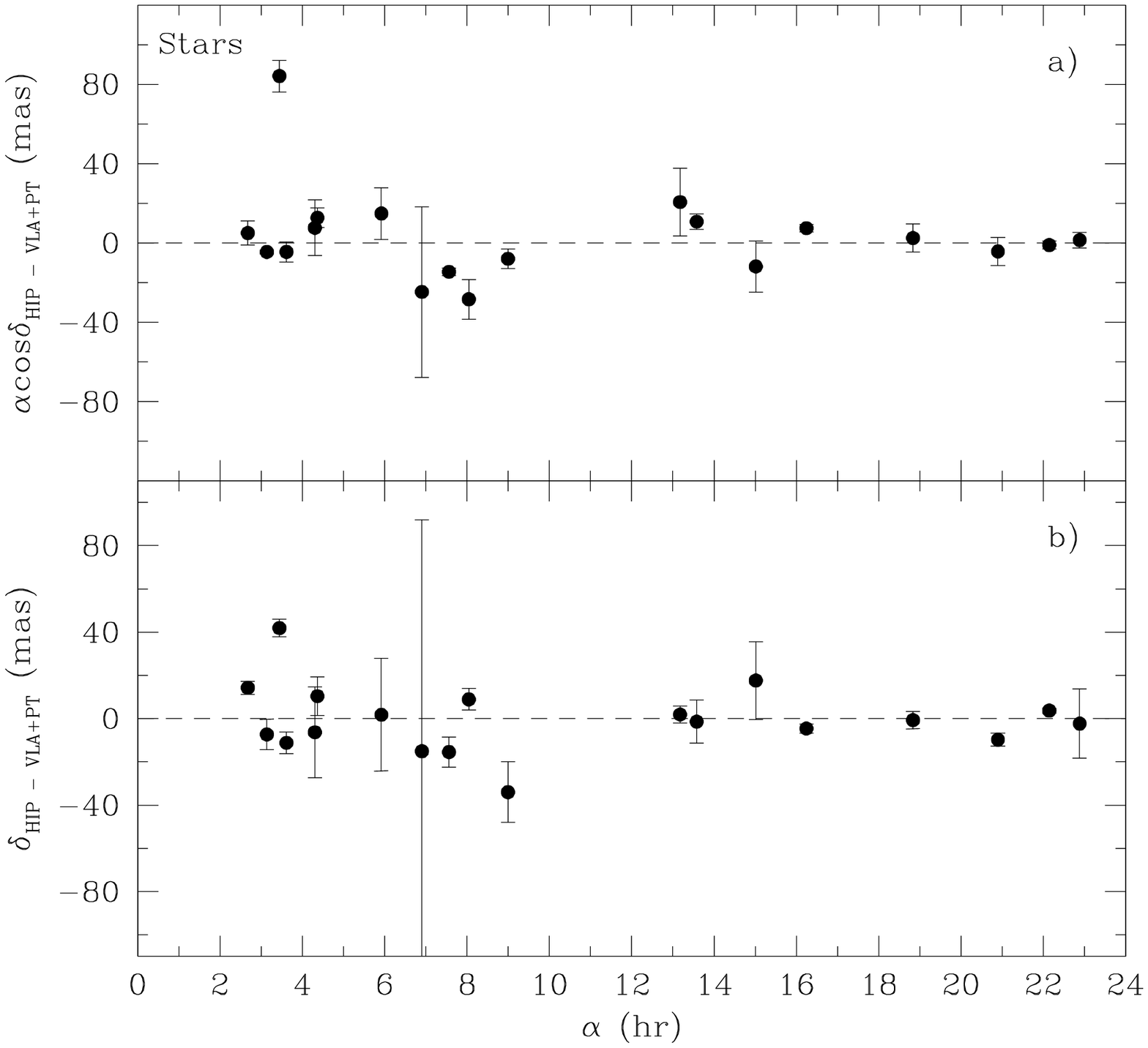}

\centerline{Figure 5}
\end{figure}

\begin{figure}[hbt]
\plotone{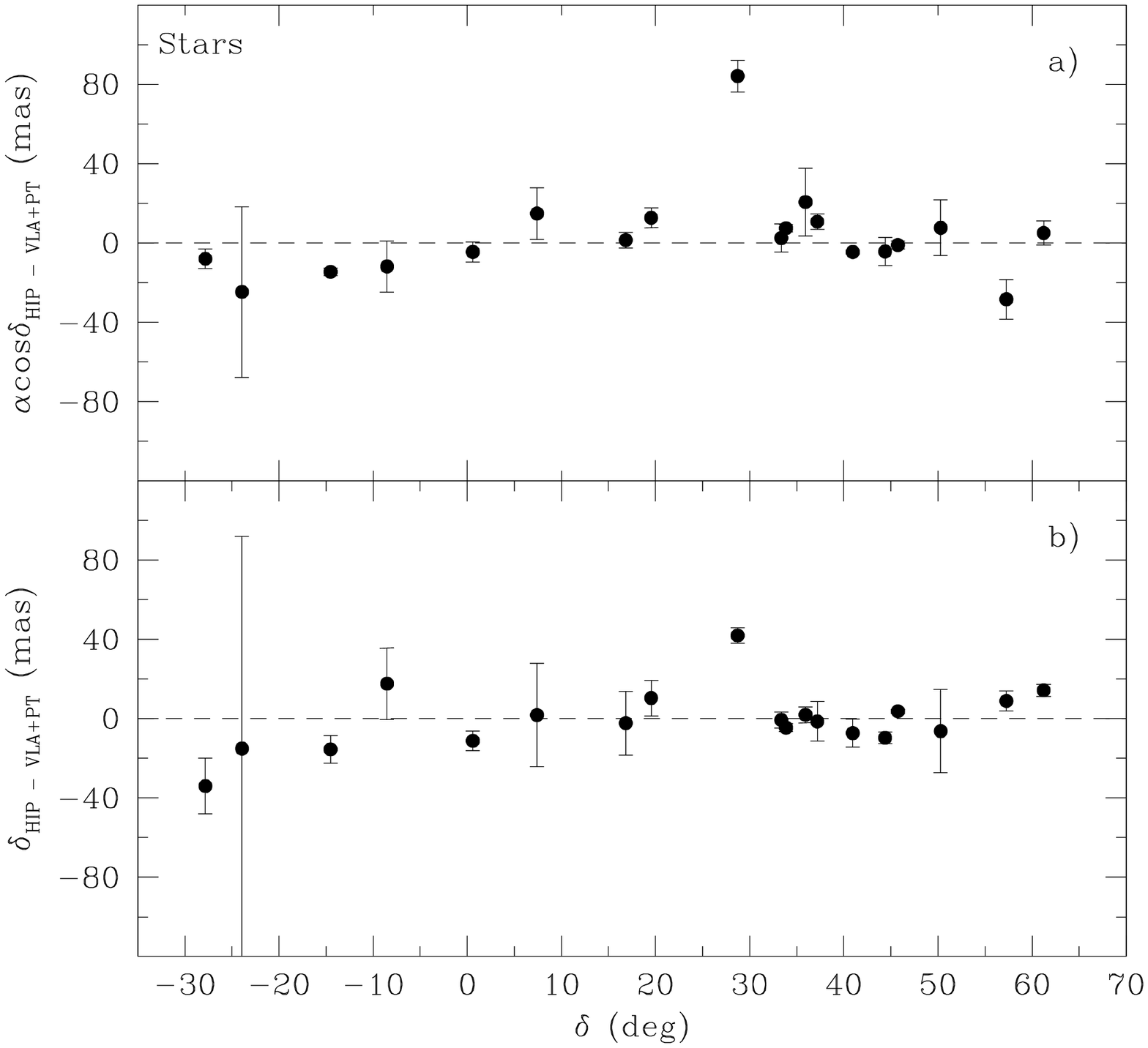}

\centerline{Figure 6}
\end{figure}

\begin{figure}[hbt]
\plotone{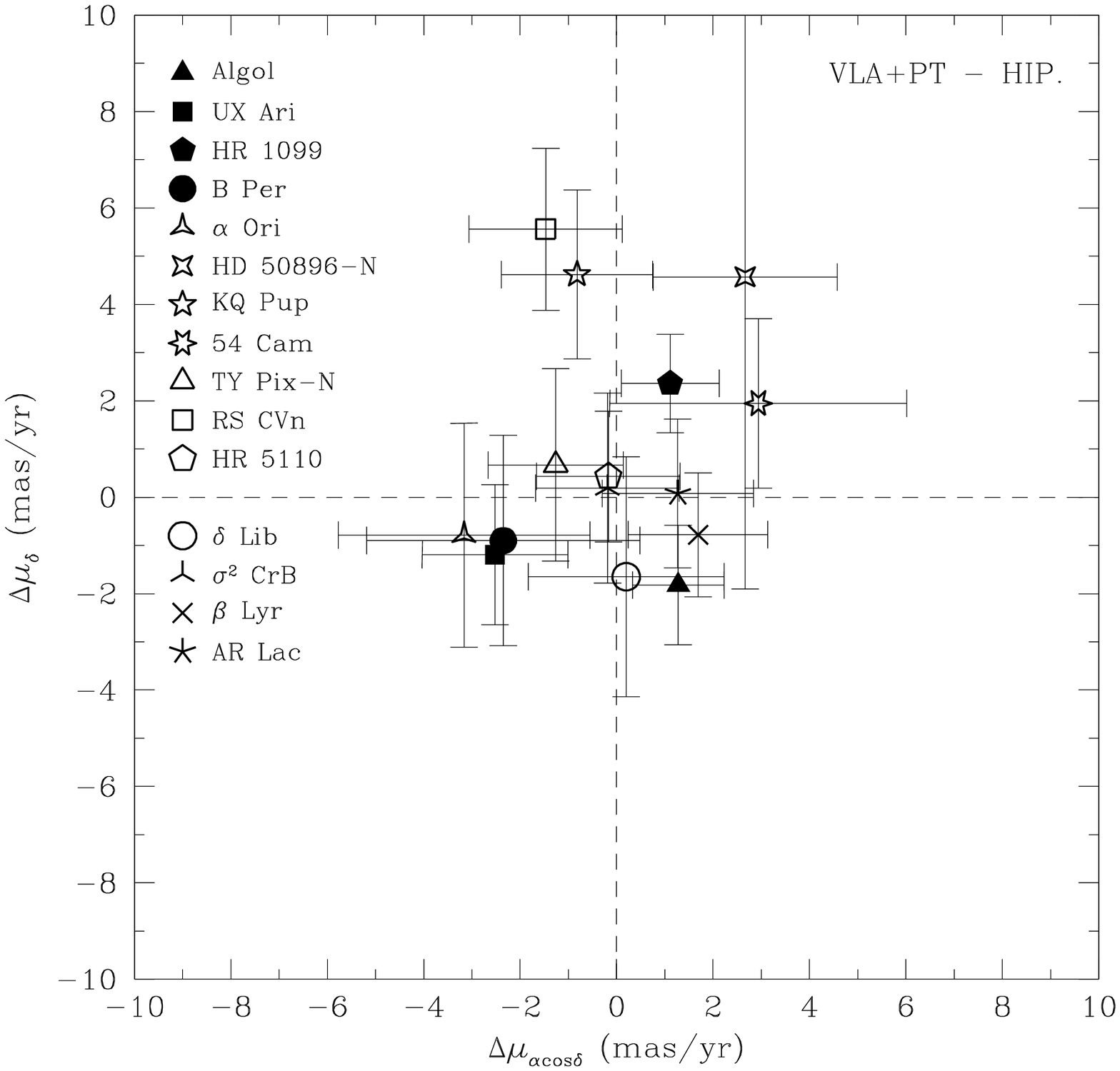}

\centerline{Figure 7}
\end{figure}

\begin{figure}[hbt]
\plotone{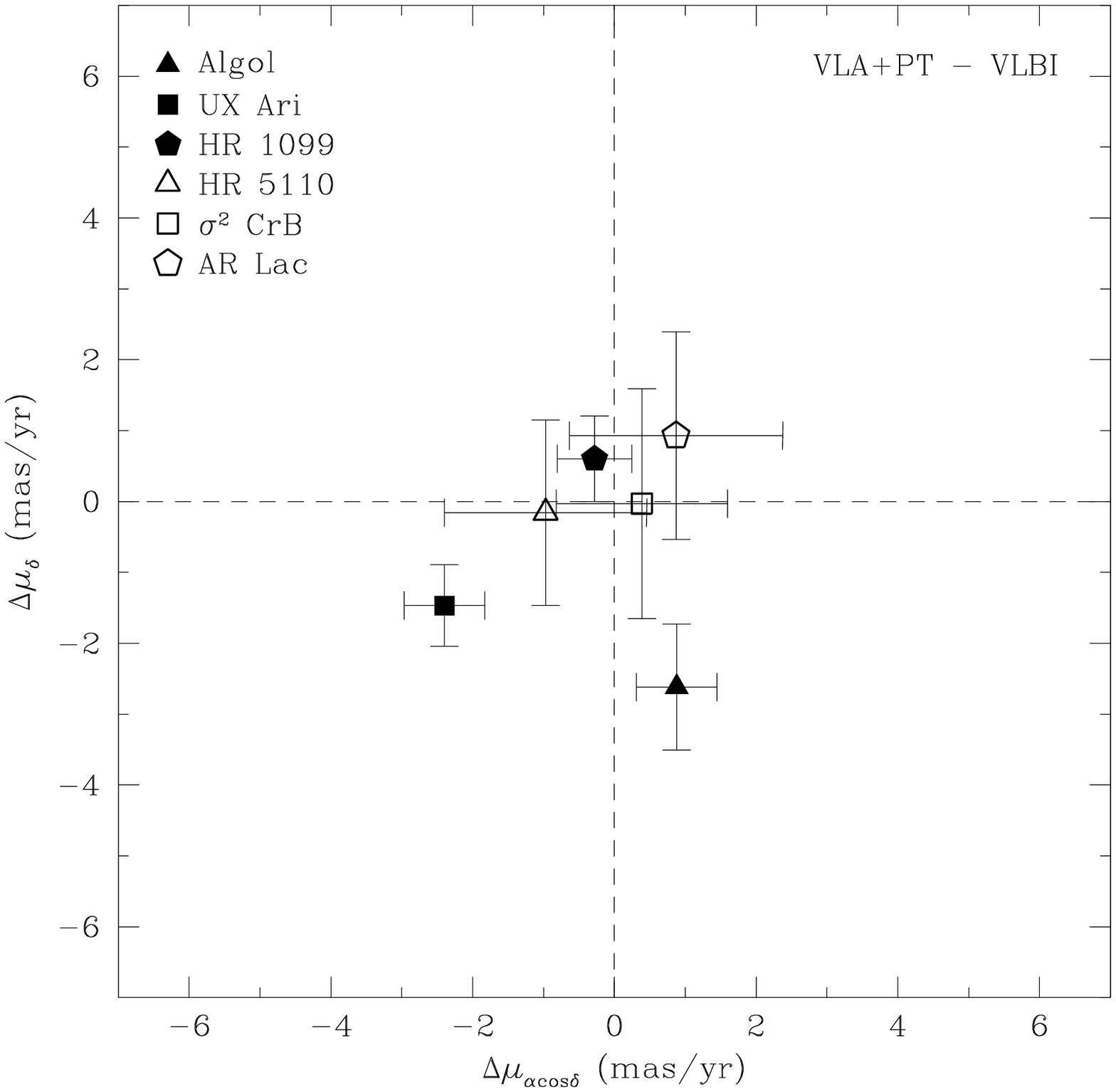}

\centerline{Figure 8}
\end{figure}

\begin{figure}[hbt]
\plotone{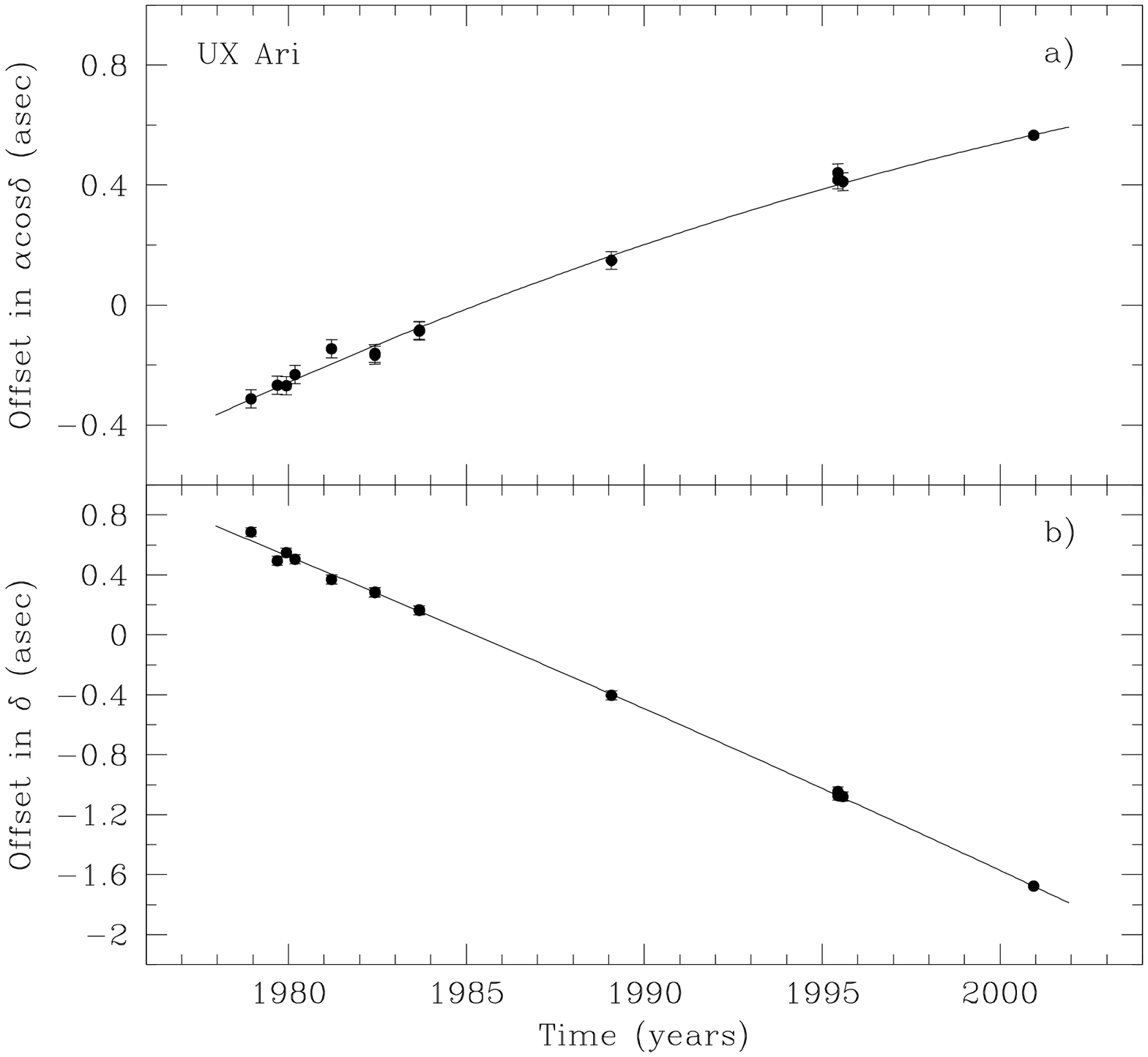}

\centerline{Figure 9}
\end{figure}

\end{document}